\documentclass[twocolumn,aps,prx,superscriptaddress]{revtex4-2}
\usepackage{graphicx}
\usepackage{amsmath}
\usepackage{amssymb}
\usepackage{braket}
\usepackage{physics}
\usepackage[colorlinks=true,linkcolor=blue,citecolor=blue,urlcolor=blue]{hyperref}

\begin{document}

\title{An Algorithm for Fast Assembling Large-Scale Defect-Free Atom Arrays}
\author{Tao Zhang}
\affiliation{Institute for Advanced Study, Tsinghua University, Beijing, 100084, China}
\affiliation{Intelligent Quantum Inception Inc., Haidian, Beijing, 100085, China} 
\affiliation{iFLYTEK Research, Hefei 230088, China}
\author{Xiaodi Li}
\affiliation{Intelligent Quantum Inception Inc., Haidian, Beijing, 100085, China} 
\affiliation{iFLYTEK Research, Hefei 230088, China}
\author{Hui Zhai}
\email{hzhai@tsinghua.edu.cn}
\affiliation{Institute for Advanced Study, Tsinghua University, Beijing, 100084, China}
\author{Linghui Chen}
\email{lhchen@iflytek.com}
\affiliation{iFLYTEK Research, Hefei 230088, China}
\affiliation{Intelligent Quantum Inception Inc., Haidian, Beijing, 100085, China} 
\date{\today}

\begin{abstract}

It is widely believed that tens of thousands of physical qubits are needed to build a practically useful quantum computer. Atom arrays formed by optical tweezers are among the most promising platforms for achieving this goal, owing to the excellent scalability and mobility of atomic qubits. However, assembling a defect-free atom array with $\sim 10^4$ qubits remains algorithmically challenging, alongside other hardware limitations. This is due to the computationally hard path-planning problems and the time-consuming generation of sufficiently smooth trajectories for optical tweezer potentials by spatial light modulators (SLM). Here, we present a unified framework comprising two innovative components to fully address these algorithmic challenges: (1) a path-planning module that employs a supervised learning approach using a graph neural network combined with a modified auction decoder, and (2) a potential-generation module called the phase and profile-aware Weighted Gerchberg-Saxton algorithm. The inference time for the first module is nearly a size-independent constant overhead of $\sim 5$ ms, and the second module generates a potential frame with about $0.5$ ms, a timescale shorter than the current commercial SLM refresh time. Altogether, our algorithm enables the assembly of an atom array with $10^4$ qubits on a timescale much shorter than the typical vacuum lifetime of the trapped atoms.

\end{abstract}

\maketitle

Atom array quantum processors are making rapid progress in qubit scalability, operational fidelity, and fully exploiting their hardware advantages of reconfigurability and high connectivity, establishing themselves as a promising platform for practical fault-tolerant quantum computing~\cite{kaufman2021quantum, bluvstein2022quantum, graham2022multi, evered2023high, bluvstein2024logical, bluvstein2026fault}. Currently, the number of controllable physical qubits in atom arrays has reached a few thousand~\cite{chiu2025continuous, lin2025ai}, while approaching tens of thousands remains challenging, although it is widely believed to be the minimum requirement for coding hundreds of logical qubits with significantly reduced error rates for a practically useful quantum computer~\cite{saffman2019quantum, beverland2022assessing, bluvstein2024logical, cain2026shor}.

In the current architecture, the atom array quantum processor begins with a defect-free array of atoms~\cite{saffman2019quantum,barredo2016atom, endres2016atom}. To assemble such an array, current technology employs an acousto-optic deflector (AOD) as mobile tweezers to move atoms~\cite{barredo2016atom, endres2016atom}. However, the AOD only allows atoms to move vertically or horizontally. These fundamental constraints, despite recent efforts to design faster rearrangement protocols~\cite{schymik2020enhanced, tian2023parallel, wang2023accelerating}, mean that the time cost $t_\text{a}$ for assembling a defect-free array must increase with the number of atoms $N$ in the array. However, given an atom vacuum time $\tau$, it is desirable to complete the assembly within $\tau/N$ to achieve a defect-free array; otherwise, significant atom loss would occur during the rearrangement. The condition $t_\text{a}(N)\ll \tau/N$ imposes a bottleneck that limits the current array size to thousands.

To overcome this challenge, one promising solution is to use a spatial light modulator (SLM) to replace the AOD and generate mobile tweezers~\cite{kim2016situ, kim2019gerchberg, lin2025ai}. In principle, the SLM allows tweezers to move along arbitrary trajectories, offering the promise of constant-time overhead independent of the number of atoms in the array. On the hardware side, however, the current refresh rate of SLMs remains low, and high-speed SLMs are under active development. On the algorithmic side, this approach also poses several unsolved challenges.

First, determining the optimal, collision-free paths for the simultaneous motion of more than a thousand atoms is a computationally challenging assignment problem. Exact algorithms, such as the Hungarian method, scale as $\mathcal{O}(N^3)$ and are too slow for real-time operation~\cite{lee2017defect}, whereas faster heuristics compromise global optimality, leading to longer trajectories that either increase the required time or increase motional heating~\cite{lin2025ai}. Second, and most critically for coherent transport, SLMs create mobile tweezers by switching between discrete holograms. If the intensity and phase of each tweezer are not sufficiently continuous across frames, atoms experience non-adiabatic changes in their trapping potential. These rapid fluctuations excite motional states, leading to heating or even atom loss, thereby degrading the fidelity of assembling a defect-free array.

More importantly, a successful algorithm should not only address these challenges but also complete the total inference time in significantly less than $\tau/N$. Without extensive optimization, the atomic vacuum lifetime $\tau$ can reach about $500$s, and the best record so far is about $1200$s with dedicated efforts to improve vacuum conditions~\cite{manetsch2025tweezer}. Therefore, we aim to assemble an atom array with tens of thousands of atoms on a timescale much shorter than $50$ms to avoid atom loss during this process.

\begin{figure}[t]
    \centering
    \includegraphics[width=0.9\columnwidth]{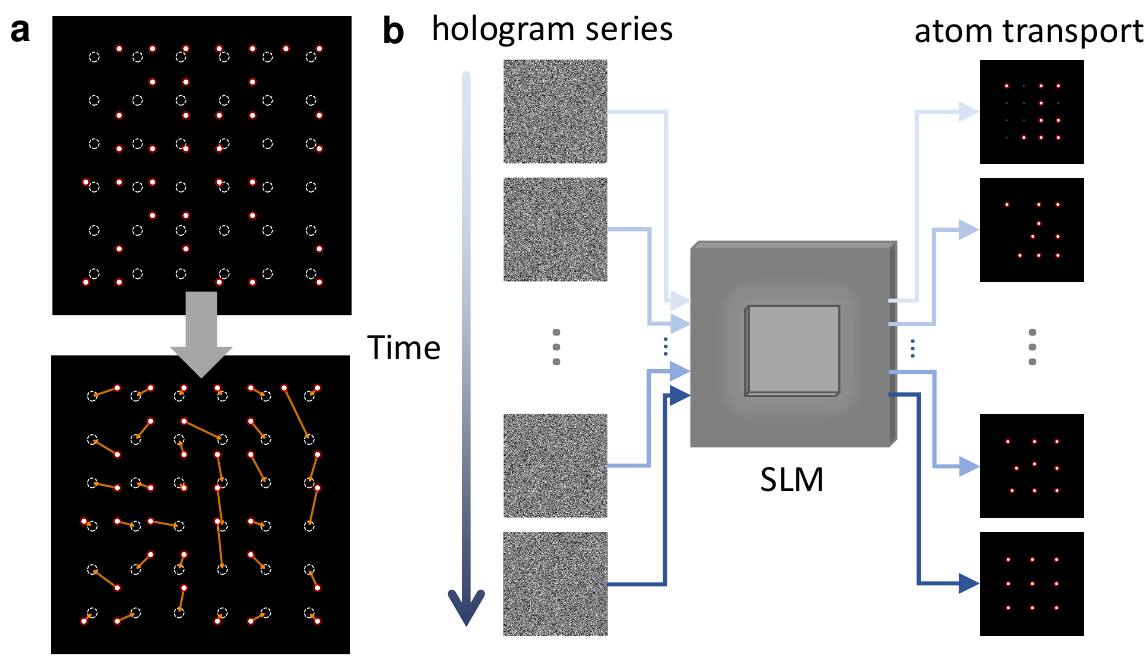}
    \caption{\textbf{Decomposition of the task.}  The entire process for assembling a large-scale atom array is divided into two stages. (a) The path planning task. The initial configuration is an array of atoms with randomly distributed vacancies. The final target configuration is a defect-free atom array. The goal of path planning is to find collision-free trajectories that simultaneously move all atoms while optimizing the maximum travel distance. (b) The potential generation task. The SLM hologram plane and the tweezer potential plane are conjugate Fourier planes. This task involves solving an inverse problem: given the planned paths, we need to compute the corresponding phase-only hologram sequence that, when applied to the input laser beams, generates the intended optical potentials and determines the required trajectories of the optical tweezers.
 }
    \label{fig:illustration}
\end{figure}

In this work, we present a unified framework that simultaneously resolves all three challenges. This framework decomposes the entire task into two synergistic tasks: path planning and potential generation, as shown in Fig.~\ref{fig:illustration}. The path-planning task is addressed by a graph neural network (GNN)~\cite{khalil2017learning}, which achieves nearly globally optimal paths with near-constant time complexity, independent of the array size. The potential generation task is addressed using an improved algorithm: we design a modified version of the Weighted Gerchberg-Saxton (WGS) algorithm~\cite{gerchberg1972practical, di2007computer}, which we name the phase and profile-aware WGS ($\text{P}^2\text{WGS}$) algorithm. Here, ``profile-aware'' refers to modeling the target tweezer amplitude as a continuous Gaussian profile rather than a single-pixel delta function, while ``phase-aware'' indicates the explicit constraint on the tweezer phase. This algorithm simultaneously optimizes both the amplitude and phase of the tweezers, thereby ensuring that the generated potential sequence is sufficiently smooth. 

\begin{figure}[t]
    \centering
    \includegraphics[width=\columnwidth]{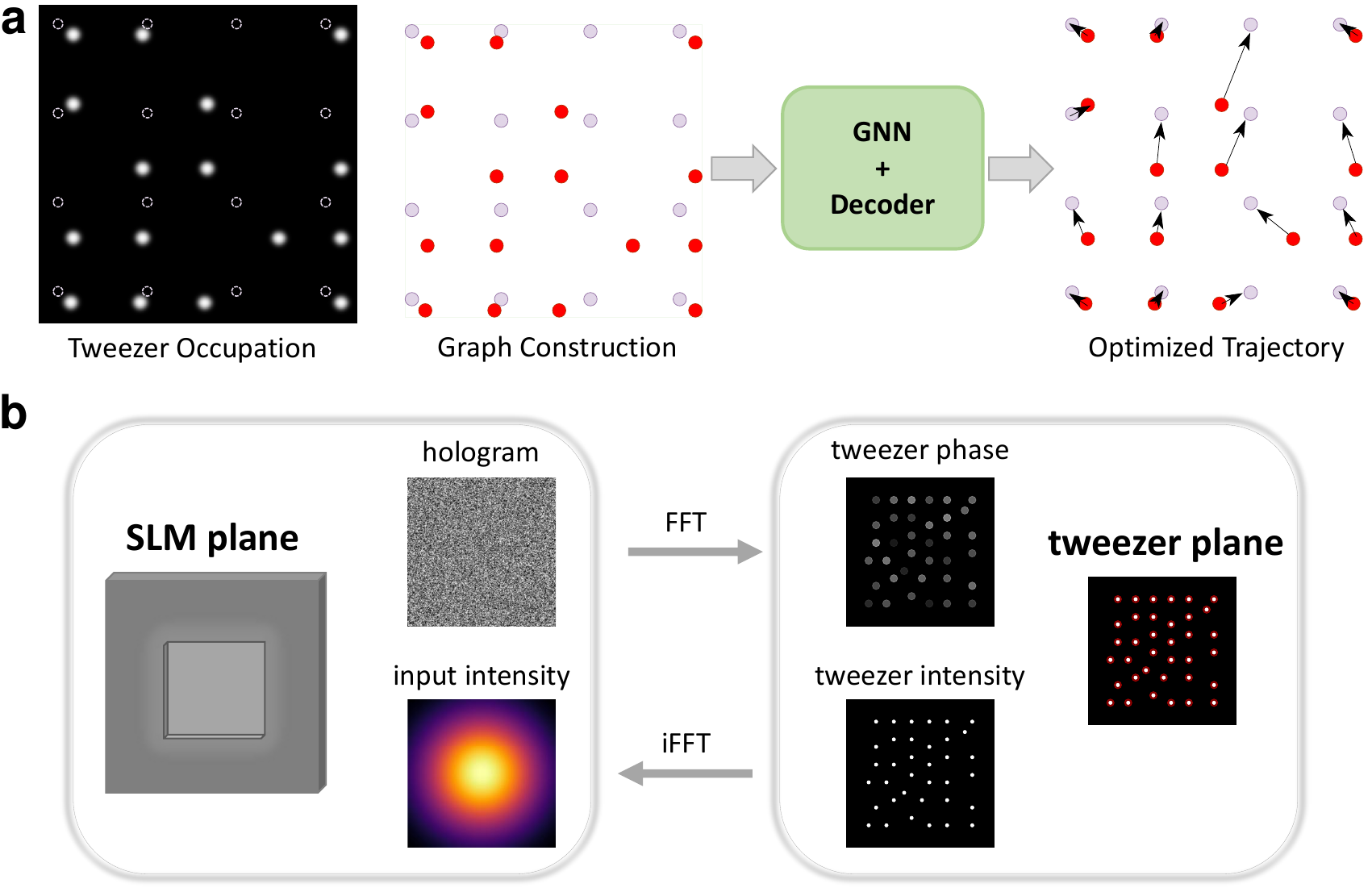}
    \caption{\textbf{Algorithm for two tasks.} 
    (a) The GNN-based path planning module. The initial and target atom sites are modeled as vertices of the graph processed by the GNN. The edges (not shown in the figure) are defined by a rule described in the main text in detail.  A parallel decoder then extracts the optimal, collision-free movement vectors from the GNN's output.
    (b) The $\text{P}^2\text{WGS}$-based potential generation module. A sequence of SLM holograms is generated via the proposed $\text{P}^2\text{WGS}$ algorithm. Unlike the traditional WGS algorithm that only constrains amplitudes, our algorithm explicitly enforces both amplitude and phase constraints in the tweezer plane during the iterations, thereby ensuring coherent and smooth transport of the atoms.}
    \label{fig:algorithm}
\end{figure}

\textit{Path Planning Task.} Initially, atoms are randomly loaded into a two-dimensional array of optical tweezers, with each tweezer having a probability $p$ of capturing a single atom and $1-p$ of being empty. Typically, $p$ ranges from $50\%$ to $80\%$. Consequently, the initial configuration is an array of atoms with randomly distributed vacancies, as represented by the red circles in Fig.~\ref{fig:algorithm}(a). The target final configuration is a defect-free array with a slightly larger interatomic distance, as indicated by the purple circles in Fig.~\ref{fig:algorithm}(a). The path planning task is to determine the one-to-one trajectories for all atoms that connect the initial and target configurations, subject to the following requirements: (i) the trajectory for every atom is a straight line; (ii) any two trajectories must avoid collision or near-collision; and (iii) the maximum and average trajectory length should be cooperatively minimized.

The Hungarian algorithm is an exact method for finding an optimal solution that satisfies these requirements ~\cite{kuhn1955hungarian, jonker1987shortest}. However, its time cost scales as $\mathcal{O}(N^3)$ and cannot afford assembling an array with even thousands of atoms. A notable example of heuristic methods is the block-wise Hungarian algorithm, which partitions the array into smaller, fixed-size independent sub-arrays~\cite{lin2025ai}. Although its runtime appears independent of array size, this approach lacks a global view because suboptimal assignments inevitably arise at block boundaries, and the total boundary length increases with array size, thereby increasing the number of suboptimal assignments. Consequently, this leads to trajectories with a maximum travel distance significantly larger than that of the global optimum.

We solve this problem with a GNN shown in Fig.~\ref{fig:algorithm}(a). To construct the input graph efficiently, we treat all tweezer sites occupied by atoms and the target sites as vertices, and attach each vertex to at most $K$ neighboring vertices ($K=128$ in practice) via edges. Each edge is of three types: atom-to-atom, target-to-target, and atom-to-target. In the input, each edge is assigned a rich feature vector that encodes its type and the squared distance. We use data generated by the globally optimal Hungarian algorithm as ground truth, in which the edge corresponding to the optimal path is assigned probability $1$, and all other edges are assigned probability $0$. During inference, the GNN outputs a probability for each candidate edge, which is subsequently processed by a fast, parallel decoder based on a modified auction algorithm~\cite{bertsekas1988auction}. Unlike greedy approaches that yield suboptimal local solutions, the auction algorithm iteratively resolves conflicts through a bidding mechanism, converging to a globally consistent, near-optimal assignment, and its inherent parallelism makes it exceptionally suitable for GPU execution. 

\begin{figure}[t]
    \centering
    \includegraphics[width=0.9\columnwidth]{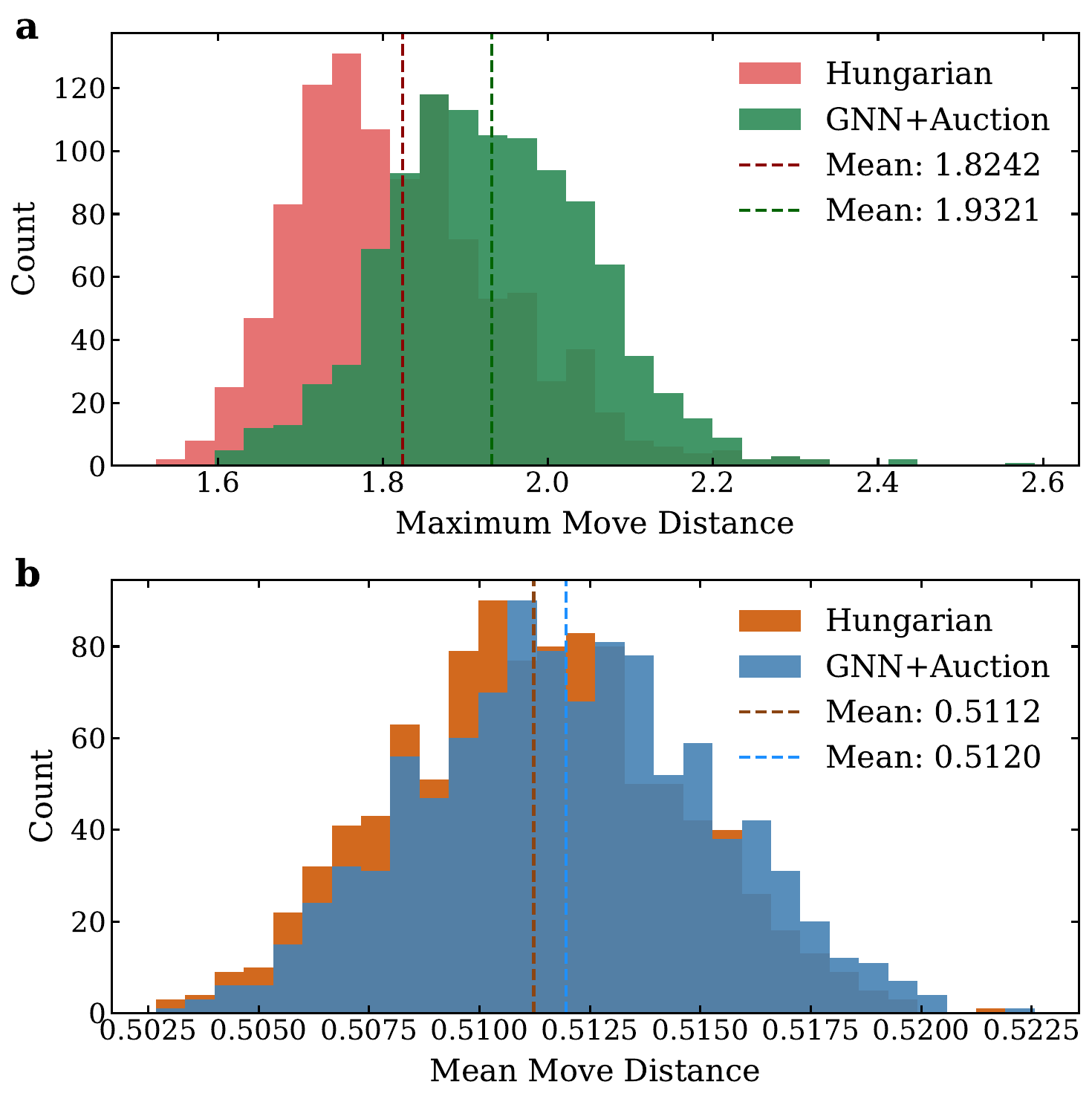}
    \caption{\textbf{GNN-based Path Planner Performance.} 
        We benchmark our method on $1024$ random instances of rearranging a stochastically loaded $127 \times 127$ initial array to a $101 \times 101$ target lattice. The initial lattice spacing is normalized to $1.0$ for simplicity. (a) Distribution of the maximum move distance. Our method (green) yields a mean of $1.93$, compared to the optimal value of $1.82$ achieved by the Hungarian algorithm (red). (b) Distribution of the average move distance. Our method (orange) achieves a mean distance of $0.5120$, which is nearly identical to the global optimum of $0.5112$ provided by the Hungarian algorithm (blue).
    }
    \label{fig:gnn_performance}
\end{figure}

For example, considering $10^4$ atoms, the 6-layer GNN was trained on a million-scale dataset for a total of 288 GPU hours using 4 NVIDIA A40 GPUs. After training, the path planning results are compared with the optimal outcomes from the Hungarian algorithm in Fig.~\ref{fig:gnn_performance}. Across $\sim 10^3$ samples, our method yields a mean maximum move distance only $6\%$ larger than the optimal results from the Hungarian algorithm, while the mean average move distance is nearly identical. Hence, we conclude that our GNN method achieves nearly optimal assignment results.

\textit{Potential Generation Task.}  Experimentally, computer-generated holograms are applied to an SLM to modulate the incident laser beam, creating the desired optical tweezers in the focal plane. The forward process, calculating the amplitude and phase of the laser in the tweezer plane from the SLM's phase hologram and the incident laser profile, is a straightforward optical Fourier transform. However, determining the required SLM phase hologram for a given target amplitude and phase distribution in the tweezer plane is a highly nontrivial inverse problem. For coherent transport of atomic qubits, solving this inverse problem must strictly satisfy two requirements: (i) the generated tweezer intensity and phase must remain sufficiently continuous across consecutive frames to prevent non-adiabatic motional heating; (ii) the computation time must be short enough to allow real-time operation within the vacuum lifetime of the array.

Traditionally, iterative approaches such as the WGS algorithm have been the standard solution to this inverse problem~\cite{gerchberg1972practical, di2007computer, pasienski2008high}. However, WGS is fundamentally designed to optimize only the target amplitude, failing to constrain the phase of the tweezers and thereby severely violating the coherence requirement for atom transport. Recently, a CNN-based approach was proposed to generate phase-continuous holograms~\cite{lin2025ai}. Although it avoids the iterative process, its inference time remains relatively long compared with the refresh rate of currently available SLMs.

To overcome these limitations, we developed the P$^2$WGS algorithm, which introduces two crucial improvements over the traditional WGS method. First, by explicitly incorporating a phase constraint in the target plane during the iterative process, our algorithm solves the inverse problem while simultaneously constraining both the intensity and phase of the optical tweezers, thereby guaranteeing coherence between consecutive frames. Second, instead of a single-pixel delta function, the target profile of each tweezer is initialized and modeled as a continuous Gaussian distribution. This physically motivated setting not only achieves sub-pixel positioning accuracy but also dramatically accelerates convergence. Owing to this rapid convergence, the number of required iterations, and thus the total time cost, can be significantly reduced. 

\begin{figure}[t]
    \centering
    \includegraphics[width=\columnwidth]{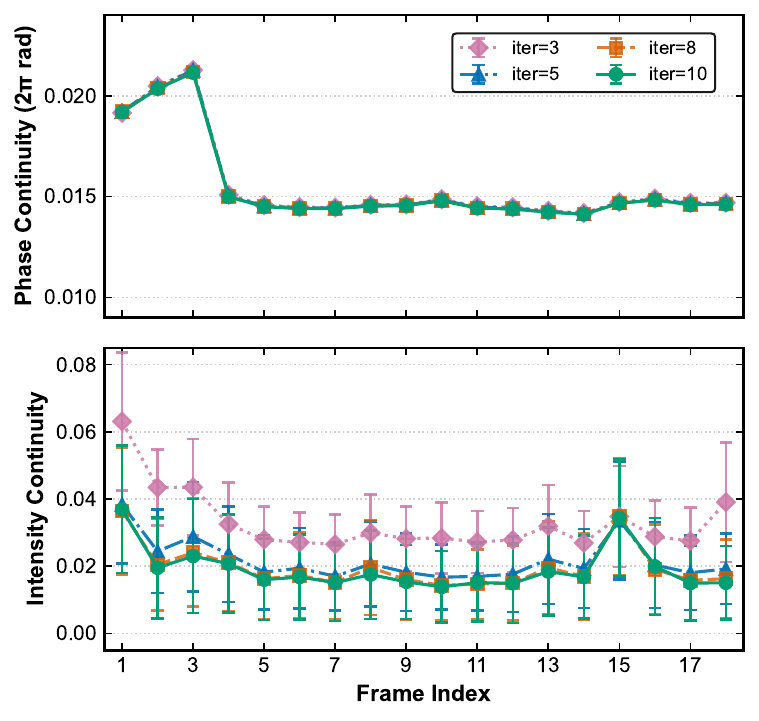}
    \caption{\textbf{Continuity of Optical Potentials.} The phase continuity (top) and intensity continuity (bottom) between consecutive frames along the atomic trajectories. The metrics are evaluated on an array assembling task for $N=10201$ ($T=101$) across different numbers of $\text{P}^2\text{WGS}$ iterations (iter = 3, 5, 8, and 10). Error bars represent the standard deviation across multiple independent stochastic loading realizations.}
    
    \label{fig:continuity}
\end{figure}

To rigorously evaluate the smoothness of the generated optical potentials, we explicitly define and track the trap properties along the atomic trajectories. Let $I_{n}^{(j)}$ denote the total intensity of the $j$-th tweezer at the $n$-th frame, computed as the sum of squared optical amplitudes within a local target neighborhood (e.g., $5 \times 5$ pixels) centered at the tweezer's position. The overall intensity continuity at the $n$-th frame is evaluated by averaging the relative absolute difference over all $N$ tweezers in the target array, $\frac{1}{N}\sum_{j=1}^N |I_{n+1}^{(j)} - I_{n}^{(j)}| / \bar{I}_{n, n+1}^{(j)}$, where $\bar{I}_{n, n+1}^{(j)}=(I_{n+1}^{(j)} + I_{n}^{(j)})/2$. Similarly, let $\phi_{n}^{(j)}$ be the optical phase evaluated at the nearest discrete pixel to the $j$-th tweezer center. The overall phase continuity is quantified by averaging the absolute phase shift $|\phi_{n+1}^{(j)} - \phi_n^{(j)}|$ wrapped to the principal range $[-\pi, \pi]$ and normalized by $2\pi$. To ensure statistical robustness against the randomness in the initial atom loading, both metrics are further averaged across multiple independent stochastic realizations. As shown in Fig.~\ref{fig:continuity}, our P$^2$WGS algorithm tightly constrains the phase fluctuations across all tested iteration numbers, fulfilling a critical requirement for coherent transport that is fundamentally ignored in the traditional WGS algorithm. To determine the optimal computational setup, we evaluate the intensity continuity under 3, 5, 8, and 10 iterations. We observe a significant improvement in intensity continuity when increasing the iterations from three to five, indicating that three iterations are insufficient for convergence. However, as the number of iterations increases to eight or ten, the intensity fluctuations remain largely unchanged, demonstrating that the algorithm has rapidly converged. Consequently, selecting five iterations offers an optimal balance between potential smoothness and computational efficiency. With only five iterations, the intensity variations are suppressed to below $4\%$ for the majority of the trajectory, typically around $2\% ~$\footnote{In practice, the initial stochastic loading typically captures more atoms than the required target array size. To assemble a defect-free array, the redundant atoms must be released. In our algorithm, before switching off the redundant optical tweezers at the fourth frame, we apply randomized spatial perturbations to their positions during the first three frames to avoid optical interference. Because the SLM hologram generates all tweezers simultaneously via global Fourier interference, the random displacement of these redundant tweezers introduces global potential perturbations, leading to the slightly larger phase and intensity fluctuations observed in these initial frames. Once the redundant tweezers are completely removed at the fourth frame, the remaining tweezers move strictly along smooth, deterministic linear trajectories, allowing the optical potential to stabilize rapidly.}. Such continuous variations guarantee the adiabaticity of the atomic transport, thereby preventing motional heating. 

\begin{figure}[t]
    \centering
    \includegraphics[width=\columnwidth]{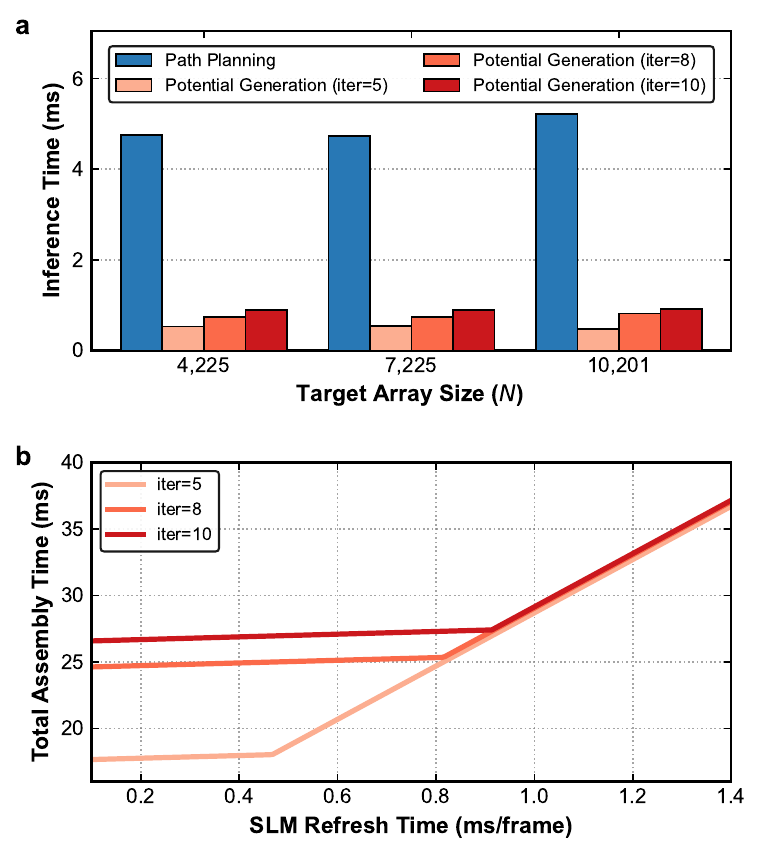}
    \caption{\textbf{Inference Time and Pipelined Assembly Performance.} (a) Computational time of the path planning and potential generation tasks across varying array sizes, evaluated on a single NVIDIA RTX 5090 GPU. The path planning (blue bars) achieves a nearly constant overhead of $\sim 5$~ms. The potential generation time (orange/red bars) remains strictly below $1$~ms per frame for $N=10,201$. (b) Total assembly time versus the hardware SLM refresh time for $N=10,201$ atoms using a 20-frame trajectory. The pipelined execution model assumes a $3$~ms constant data transfer delay. A kink occurs when the SLM refresh time matches the algorithm's per-frame generation time. To the left of this kink, the system is compute-bound (plateau region), whereas to the right, the algorithm overhead is covered, rendering the total time strictly hardware-bound (linear region).}
    \label{fig:time}
\end{figure}

\textit{Inference Time.}  Aside from generating high-fidelity potentials, our framework is highly optimized for real-time execution. In Fig.~\ref{fig:time}(a), we compare the computational time of the path planning module and the potential generation module across varying target array sizes $N$. Benefiting from the fully parallelized graph construction and the auction decoder, the GNN-based path planner exhibits a nearly constant $\mathcal{O}(1)$ time complexity, finding globally near-optimal assignments in approximately $5$~ms, even for systems exceeding $10^4$ atoms. Meanwhile, the P$^2$WGS algorithm handles hologram generation with remarkable efficiency. Based on the convergence analysis above, we select five iterations as the standard setup for practical implementation. Under this configuration, evaluating a single frame for $N \approx 10^4$ atoms takes around $0.5$~ms on a single commercial GPU. In a practical experimental setup, the hologram generation and the physical SLM refresh can be executed in a pipelined, parallel manner.

To systematically evaluate this, in Fig.~\ref{fig:time}(b) we model the total assembly time as a function of the SLM refresh time for a 20-frame trajectory, incorporating a constant hardware data transfer delay assumed to be $3$~ms. This pipelined execution reveals two distinct performance regimes. If the algorithmic generation time exceeds the SLM refresh cycle, the hardware stalls waiting for the holograms, placing the system in a compute-bound regime characterized by the plateau on the left, with a total assembly time of $\sim 18.5$~ms for 5 iterations. Conversely, if the algorithmic generation time is shorter than the SLM refresh cycle, the computational module can continuously stream holograms ahead of the hardware display, placing the system in a hardware-bound regime characterized by the linear regime on the right. Currently, the typical refresh cycle of commercial high-speed SLMs for generating large-scale tweezer arrays is around $\sim 1$~ms, which is longer than our per-frame inference time ($\sim 0.5$~ms for 5 iterations); therefore, the total time cost for assembling a defect-free array of $10^4$ atoms is strictly dominated by the physical hardware limit (e.g., $\sim 28.5$~ms for a $1$-ms refresh rate SLM). Meanwhile, the speed of SLMs is advancing rapidly~\cite{wei202610, bytyqi2026device}, which might push performance back into a compute-bound regime in the near future.

\textit{Conclusion.} We have developed a highly-efficient framework that achieves global optimality in path planning and generates sufficiently smooth tweezer potentials along the trajectories for assembling a defect-free array of atomic qubits, with total inference time well within the vacuum lifetime even for tens of thousands of atomic qubits. Reconfigurability is one of the major advantages of the atom array platform for realizing fault-tolerant quantum computation. Although we use the task of assembling the initial atom array to illustrate the power of our algorithm, it can be directly generalized to other tasks that require dynamic atom rearrangement, such as performing non-local parallel gate operations~\cite{evered2023high, bluvstein2024logical, bluvstein2026fault} or encoding quantum low-density parity-check codes~\cite{breuckmann2021quantum, xu2024constant}. Based on this algorithm, we have developed a software package called ``Zhuifeng''~\footnote{``Zhuifeng,'' which literally translates to ``Chasing the Wind,'' was the fastest and most beloved horse of Emperor Qin Shi Huang, the first emperor of the Qin Dynasty.} that enables rapid movement of atomic qubits using  SLM only. The current version of this software is already sufficient for systems operating at the scale of tens of thousands of qubits. This breakthrough provides an essential tool for early fault-tolerant quantum computation on the atom array platform.

\begin{acknowledgments}
    This work is supported by the National Natural Science Foundation of China under Grant No.~12488301 (H.Z.) and No.~U23A6004 (H.Z.).
\end{acknowledgments}
    
\bibliography{references.bib}

\begin{thebibliography}{33}%
\makeatletter
\providecommand \@ifxundefined [1]{%
 \@ifx{#1\undefined}
}%
\providecommand \@ifnum [1]{%
 \ifnum #1\expandafter \@firstoftwo
 \else \expandafter \@secondoftwo
 \fi
}%
\providecommand \@ifx [1]{%
 \ifx #1\expandafter \@firstoftwo
 \else \expandafter \@secondoftwo
 \fi
}%
\providecommand \natexlab [1]{#1}%
\providecommand \enquote  [1]{``#1''}%
\providecommand \bibnamefont  [1]{#1}%
\providecommand \bibfnamefont [1]{#1}%
\providecommand \citenamefont [1]{#1}%
\providecommand \href@noop [0]{\@secondoftwo}%
\providecommand \href [0]{\begingroup \@sanitize@url \@href}%
\providecommand \@href[1]{\@@startlink{#1}\@@href}%
\providecommand \@@href[1]{\endgroup#1\@@endlink}%
\providecommand \@sanitize@url [0]{\catcode `\\12\catcode `\$12\catcode `\&12\catcode `\#12\catcode `\^12\catcode `\_12\catcode `\%12\relax}%
\providecommand \@@startlink[1]{}%
\providecommand \@@endlink[0]{}%
\providecommand \url  [0]{\begingroup\@sanitize@url \@url }%
\providecommand \@url [1]{\endgroup\@href {#1}{\urlprefix }}%
\providecommand \urlprefix  [0]{URL }%
\providecommand \Eprint [0]{\href }%
\providecommand \doibase [0]{https://doi.org/}%
\providecommand \selectlanguage [0]{\@gobble}%
\providecommand \bibinfo  [0]{\@secondoftwo}%
\providecommand \bibfield  [0]{\@secondoftwo}%
\providecommand \translation [1]{[#1]}%
\providecommand \BibitemOpen [0]{}%
\providecommand \bibitemStop [0]{}%
\providecommand \bibitemNoStop [0]{.\EOS\space}%
\providecommand \EOS [0]{\spacefactor3000\relax}%
\providecommand \BibitemShut  [1]{\csname bibitem#1\endcsname}%
\let\auto@bib@innerbib\@empty
\bibitem [{\citenamefont {Kaufman}\ and\ \citenamefont {Ni}(2021)}]{kaufman2021quantum}%
  \BibitemOpen
  \bibfield  {author} {\bibinfo {author} {\bibfnamefont {A.~M.}\ \bibnamefont {Kaufman}}\ and\ \bibinfo {author} {\bibfnamefont {K.-K.}\ \bibnamefont {Ni}},\ }\bibfield  {title} {\bibinfo {title} {Quantum science with optical tweezer arrays of ultracold atoms and molecules},\ }\href@noop {} {\bibfield  {journal} {\bibinfo  {journal} {Nature Physics}\ }\textbf {\bibinfo {volume} {17}},\ \bibinfo {pages} {1324} (\bibinfo {year} {2021})}\BibitemShut {NoStop}%
\bibitem [{\citenamefont {Bluvstein}\ \emph {et~al.}(2022)\citenamefont {Bluvstein}, \citenamefont {Levine}, \citenamefont {Semeghini}, \citenamefont {Wang}, \citenamefont {Ebadi}, \citenamefont {Kalinowski}, \citenamefont {Keesling}, \citenamefont {Maskara}, \citenamefont {Pichler}, \citenamefont {Greiner}, \citenamefont {Vuletić},\ and\ \citenamefont {Lukin}}]{bluvstein2022quantum}%
  \BibitemOpen
  \bibfield  {author} {\bibinfo {author} {\bibfnamefont {D.}~\bibnamefont {Bluvstein}}, \bibinfo {author} {\bibfnamefont {H.}~\bibnamefont {Levine}}, \bibinfo {author} {\bibfnamefont {G.}~\bibnamefont {Semeghini}}, \bibinfo {author} {\bibfnamefont {T.~T.}\ \bibnamefont {Wang}}, \bibinfo {author} {\bibfnamefont {S.}~\bibnamefont {Ebadi}}, \bibinfo {author} {\bibfnamefont {M.}~\bibnamefont {Kalinowski}}, \bibinfo {author} {\bibfnamefont {A.}~\bibnamefont {Keesling}}, \bibinfo {author} {\bibfnamefont {N.}~\bibnamefont {Maskara}}, \bibinfo {author} {\bibfnamefont {H.}~\bibnamefont {Pichler}}, \bibinfo {author} {\bibfnamefont {M.}~\bibnamefont {Greiner}}, \bibinfo {author} {\bibfnamefont {V.}~\bibnamefont {Vuletić}},\ and\ \bibinfo {author} {\bibfnamefont {M.~D.}\ \bibnamefont {Lukin}},\ }\bibfield  {title} {\bibinfo {title} {A quantum processor based on coherent transport of entangled atom arrays},\ }\href@noop {} {\bibfield  {journal} {\bibinfo  {journal} {Nature}\ }\textbf {\bibinfo {volume} {604}},\ \bibinfo {pages} {451} (\bibinfo {year} {2022})}\BibitemShut {NoStop}%
\bibitem [{\citenamefont {Graham}\ \emph {et~al.}(2022)\citenamefont {Graham}, \citenamefont {Song}, \citenamefont {Scott}, \citenamefont {Poole}, \citenamefont {Phuttitarn}, \citenamefont {Jooya}, \citenamefont {Eichler}, \citenamefont {Jiang}, \citenamefont {Marra}, \citenamefont {Grinkemeyer}, \citenamefont {Kwon}, \citenamefont {Ebert}, \citenamefont {Cherek}, \citenamefont {Lichtman}, \citenamefont {Gillette}, \citenamefont {Gilbert}, \citenamefont {Bowman}, \citenamefont {Ballance}, \citenamefont {Campbell}, \citenamefont {Dahl}, \citenamefont {Crawford}, \citenamefont {Blunt}, \citenamefont {Rogers}, \citenamefont {Noel},\ and\ \citenamefont {Saffman}}]{graham2022multi}%
  \BibitemOpen
  \bibfield  {author} {\bibinfo {author} {\bibfnamefont {T.~M.}\ \bibnamefont {Graham}}, \bibinfo {author} {\bibfnamefont {Y.}~\bibnamefont {Song}}, \bibinfo {author} {\bibfnamefont {J.}~\bibnamefont {Scott}}, \bibinfo {author} {\bibfnamefont {C.}~\bibnamefont {Poole}}, \bibinfo {author} {\bibfnamefont {L.}~\bibnamefont {Phuttitarn}}, \bibinfo {author} {\bibfnamefont {K.}~\bibnamefont {Jooya}}, \bibinfo {author} {\bibfnamefont {P.}~\bibnamefont {Eichler}}, \bibinfo {author} {\bibfnamefont {X.}~\bibnamefont {Jiang}}, \bibinfo {author} {\bibfnamefont {A.}~\bibnamefont {Marra}}, \bibinfo {author} {\bibfnamefont {B.}~\bibnamefont {Grinkemeyer}}, \bibinfo {author} {\bibfnamefont {M.}~\bibnamefont {Kwon}}, \bibinfo {author} {\bibfnamefont {M.}~\bibnamefont {Ebert}}, \bibinfo {author} {\bibfnamefont {J.}~\bibnamefont {Cherek}}, \bibinfo {author} {\bibfnamefont {M.~T.}\ \bibnamefont {Lichtman}}, \bibinfo {author} {\bibfnamefont {M.}~\bibnamefont {Gillette}}, \bibinfo {author} {\bibfnamefont {J.}~\bibnamefont {Gilbert}}, \bibinfo {author} {\bibfnamefont {D.}~\bibnamefont {Bowman}}, \bibinfo {author} {\bibfnamefont {T.}~\bibnamefont {Ballance}}, \bibinfo {author} {\bibfnamefont {C.}~\bibnamefont {Campbell}}, \bibinfo {author} {\bibfnamefont {E.~D.}\ \bibnamefont {Dahl}}, \bibinfo {author} {\bibfnamefont {O.}~\bibnamefont {Crawford}}, \bibinfo {author} {\bibfnamefont {N.~S.}\ \bibnamefont {Blunt}}, \bibinfo {author} {\bibfnamefont {B.}~\bibnamefont {Rogers}}, \bibinfo {author} {\bibfnamefont {T.}~\bibnamefont {Noel}},\ and\ \bibinfo {author} {\bibfnamefont {M.}~\bibnamefont {Saffman}},\ }\bibfield  {title} {\bibinfo {title} {Multi-qubit entanglement and algorithms on a neutral-atom quantum computer},\ }\href@noop {} {\bibfield  {journal} {\bibinfo  {journal} {Nature}\ }\textbf {\bibinfo {volume} {604}},\ \bibinfo {pages} {457} (\bibinfo {year} {2022})}\BibitemShut {NoStop}%
\bibitem [{\citenamefont {Evered}\ \emph {et~al.}(2023)\citenamefont {Evered}, \citenamefont {Bluvstein}, \citenamefont {Kalinowski}, \citenamefont {Ebadi}, \citenamefont {Manovitz}, \citenamefont {Zhou}, \citenamefont {Li}, \citenamefont {Geim}, \citenamefont {Wang}, \citenamefont {Maskara}, \citenamefont {Levine}, \citenamefont {Semeghini}, \citenamefont {Greiner}, \citenamefont {Vuletić},\ and\ \citenamefont {Lukin}}]{evered2023high}%
  \BibitemOpen
  \bibfield  {author} {\bibinfo {author} {\bibfnamefont {S.~J.}\ \bibnamefont {Evered}}, \bibinfo {author} {\bibfnamefont {D.}~\bibnamefont {Bluvstein}}, \bibinfo {author} {\bibfnamefont {M.}~\bibnamefont {Kalinowski}}, \bibinfo {author} {\bibfnamefont {S.}~\bibnamefont {Ebadi}}, \bibinfo {author} {\bibfnamefont {T.}~\bibnamefont {Manovitz}}, \bibinfo {author} {\bibfnamefont {H.}~\bibnamefont {Zhou}}, \bibinfo {author} {\bibfnamefont {S.~H.}\ \bibnamefont {Li}}, \bibinfo {author} {\bibfnamefont {A.~A.}\ \bibnamefont {Geim}}, \bibinfo {author} {\bibfnamefont {T.~T.}\ \bibnamefont {Wang}}, \bibinfo {author} {\bibfnamefont {N.}~\bibnamefont {Maskara}}, \bibinfo {author} {\bibfnamefont {H.}~\bibnamefont {Levine}}, \bibinfo {author} {\bibfnamefont {G.}~\bibnamefont {Semeghini}}, \bibinfo {author} {\bibfnamefont {M.}~\bibnamefont {Greiner}}, \bibinfo {author} {\bibfnamefont {V.}~\bibnamefont {Vuletić}},\ and\ \bibinfo {author} {\bibfnamefont {M.~D.}\ \bibnamefont {Lukin}},\ }\bibfield  {title} {\bibinfo {title} {High-fidelity parallel entangling gates on a neutral-atom quantum computer},\ }\href@noop {} {\bibfield  {journal} {\bibinfo  {journal} {Nature}\ }\textbf {\bibinfo {volume} {622}},\ \bibinfo {pages} {268} (\bibinfo {year} {2023})}\BibitemShut {NoStop}%
\bibitem [{\citenamefont {Bluvstein}\ \emph {et~al.}(2024)\citenamefont {Bluvstein}, \citenamefont {Evered}, \citenamefont {Geim}, \citenamefont {Li}, \citenamefont {Zhou}, \citenamefont {Manovitz}, \citenamefont {Ebadi}, \citenamefont {Cain}, \citenamefont {Kalinowski}, \citenamefont {Hangleiter}, \citenamefont {Bonilla~Ataides}, \citenamefont {Maskara}, \citenamefont {Cong}, \citenamefont {Gao}, \citenamefont {Sales~Rodriguez}, \citenamefont {Karolyshyn}, \citenamefont {Semeghini}, \citenamefont {Gullans}, \citenamefont {Greiner}, \citenamefont {Vuletić},\ and\ \citenamefont {Lukin}}]{bluvstein2024logical}%
  \BibitemOpen
  \bibfield  {author} {\bibinfo {author} {\bibfnamefont {D.}~\bibnamefont {Bluvstein}}, \bibinfo {author} {\bibfnamefont {S.~J.}\ \bibnamefont {Evered}}, \bibinfo {author} {\bibfnamefont {A.~A.}\ \bibnamefont {Geim}}, \bibinfo {author} {\bibfnamefont {S.~H.}\ \bibnamefont {Li}}, \bibinfo {author} {\bibfnamefont {H.}~\bibnamefont {Zhou}}, \bibinfo {author} {\bibfnamefont {T.}~\bibnamefont {Manovitz}}, \bibinfo {author} {\bibfnamefont {S.}~\bibnamefont {Ebadi}}, \bibinfo {author} {\bibfnamefont {M.}~\bibnamefont {Cain}}, \bibinfo {author} {\bibfnamefont {M.}~\bibnamefont {Kalinowski}}, \bibinfo {author} {\bibfnamefont {D.}~\bibnamefont {Hangleiter}}, \bibinfo {author} {\bibfnamefont {J.~P.}\ \bibnamefont {Bonilla~Ataides}}, \bibinfo {author} {\bibfnamefont {N.}~\bibnamefont {Maskara}}, \bibinfo {author} {\bibfnamefont {I.}~\bibnamefont {Cong}}, \bibinfo {author} {\bibfnamefont {X.}~\bibnamefont {Gao}}, \bibinfo {author} {\bibfnamefont {P.}~\bibnamefont {Sales~Rodriguez}}, \bibinfo {author} {\bibfnamefont {T.}~\bibnamefont {Karolyshyn}}, \bibinfo {author} {\bibfnamefont {G.}~\bibnamefont {Semeghini}}, \bibinfo {author} {\bibfnamefont {M.~J.}\ \bibnamefont {Gullans}}, \bibinfo {author} {\bibfnamefont {M.}~\bibnamefont {Greiner}}, \bibinfo {author} {\bibfnamefont {V.}~\bibnamefont {Vuletić}},\ and\ \bibinfo {author} {\bibfnamefont {M.~D.}\ \bibnamefont {Lukin}},\ }\bibfield  {title} {\bibinfo {title} {Logical quantum processor based on reconfigurable atom arrays},\ }\href@noop {} {\bibfield  {journal} {\bibinfo  {journal} {Nature}\ }\textbf {\bibinfo {volume} {626}},\ \bibinfo {pages} {58} (\bibinfo {year} {2024})}\BibitemShut {NoStop}%
\bibitem [{\citenamefont {Bluvstein}\ \emph {et~al.}(2026)\citenamefont {Bluvstein}, \citenamefont {Geim}, \citenamefont {Li}, \citenamefont {Evered}, \citenamefont {Bonilla~Ataides}, \citenamefont {Baranes}, \citenamefont {Gu}, \citenamefont {Manovitz}, \citenamefont {Xu}, \citenamefont {Kalinowski}, \citenamefont {Majidy}, \citenamefont {Kokail}, \citenamefont {Maskara}, \citenamefont {Trapp}, \citenamefont {Stewart}, \citenamefont {Hollerith}, \citenamefont {Zhou}, \citenamefont {Gullans}, \citenamefont {Yelin}, \citenamefont {Greiner}, \citenamefont {Vuletić}, \citenamefont {Cain},\ and\ \citenamefont {Lukin}}]{bluvstein2026fault}%
  \BibitemOpen
  \bibfield  {author} {\bibinfo {author} {\bibfnamefont {D.}~\bibnamefont {Bluvstein}}, \bibinfo {author} {\bibfnamefont {A.~A.}\ \bibnamefont {Geim}}, \bibinfo {author} {\bibfnamefont {S.~H.}\ \bibnamefont {Li}}, \bibinfo {author} {\bibfnamefont {S.~J.}\ \bibnamefont {Evered}}, \bibinfo {author} {\bibfnamefont {J.~P.}\ \bibnamefont {Bonilla~Ataides}}, \bibinfo {author} {\bibfnamefont {G.}~\bibnamefont {Baranes}}, \bibinfo {author} {\bibfnamefont {A.}~\bibnamefont {Gu}}, \bibinfo {author} {\bibfnamefont {T.}~\bibnamefont {Manovitz}}, \bibinfo {author} {\bibfnamefont {M.}~\bibnamefont {Xu}}, \bibinfo {author} {\bibfnamefont {M.}~\bibnamefont {Kalinowski}}, \bibinfo {author} {\bibfnamefont {S.}~\bibnamefont {Majidy}}, \bibinfo {author} {\bibfnamefont {C.}~\bibnamefont {Kokail}}, \bibinfo {author} {\bibfnamefont {N.}~\bibnamefont {Maskara}}, \bibinfo {author} {\bibfnamefont {E.~C.}\ \bibnamefont {Trapp}}, \bibinfo {author} {\bibfnamefont {L.~M.}\ \bibnamefont {Stewart}}, \bibinfo {author} {\bibfnamefont {S.}~\bibnamefont {Hollerith}}, \bibinfo {author} {\bibfnamefont {H.}~\bibnamefont {Zhou}}, \bibinfo {author} {\bibfnamefont {M.~J.}\ \bibnamefont {Gullans}}, \bibinfo {author} {\bibfnamefont {S.~F.}\ \bibnamefont {Yelin}}, \bibinfo {author} {\bibfnamefont {M.}~\bibnamefont {Greiner}}, \bibinfo {author} {\bibfnamefont {V.}~\bibnamefont {Vuletić}}, \bibinfo {author} {\bibfnamefont {M.}~\bibnamefont {Cain}},\ and\ \bibinfo {author} {\bibfnamefont {M.~D.}\ \bibnamefont {Lukin}},\ }\bibfield  {title} {\bibinfo {title} {A fault-tolerant neutral-atom architecture for universal quantum computation},\ }\href@noop {} {\bibfield  {journal} {\bibinfo  {journal} {Nature}\ }\textbf {\bibinfo {volume} {649}},\ \bibinfo {pages} {39} (\bibinfo {year} {2026})}\BibitemShut {NoStop}%
\bibitem [{\citenamefont {Chiu}\ \emph {et~al.}(2025)\citenamefont {Chiu}, \citenamefont {Trapp}, \citenamefont {Guo}, \citenamefont {Abobeih}, \citenamefont {Stewart}, \citenamefont {Hollerith}, \citenamefont {Stroganov}, \citenamefont {Kalinowski}, \citenamefont {Geim}, \citenamefont {Evered}, \citenamefont {Li}, \citenamefont {Lyu}, \citenamefont {Peters}, \citenamefont {Bluvstein}, \citenamefont {Wang}, \citenamefont {Greiner}, \citenamefont {Vuletić},\ and\ \citenamefont {Lukin}}]{chiu2025continuous}%
  \BibitemOpen
  \bibfield  {author} {\bibinfo {author} {\bibfnamefont {N.-C.}\ \bibnamefont {Chiu}}, \bibinfo {author} {\bibfnamefont {E.~C.}\ \bibnamefont {Trapp}}, \bibinfo {author} {\bibfnamefont {J.}~\bibnamefont {Guo}}, \bibinfo {author} {\bibfnamefont {M.~H.}\ \bibnamefont {Abobeih}}, \bibinfo {author} {\bibfnamefont {L.~M.}\ \bibnamefont {Stewart}}, \bibinfo {author} {\bibfnamefont {S.}~\bibnamefont {Hollerith}}, \bibinfo {author} {\bibfnamefont {P.~L.}\ \bibnamefont {Stroganov}}, \bibinfo {author} {\bibfnamefont {M.}~\bibnamefont {Kalinowski}}, \bibinfo {author} {\bibfnamefont {A.~A.}\ \bibnamefont {Geim}}, \bibinfo {author} {\bibfnamefont {S.~J.}\ \bibnamefont {Evered}}, \bibinfo {author} {\bibfnamefont {S.~H.}\ \bibnamefont {Li}}, \bibinfo {author} {\bibfnamefont {X.}~\bibnamefont {Lyu}}, \bibinfo {author} {\bibfnamefont {L.~M.}\ \bibnamefont {Peters}}, \bibinfo {author} {\bibfnamefont {D.}~\bibnamefont {Bluvstein}}, \bibinfo {author} {\bibfnamefont {T.~T.}\ \bibnamefont {Wang}}, \bibinfo {author} {\bibfnamefont {M.}~\bibnamefont {Greiner}}, \bibinfo {author} {\bibfnamefont {V.}~\bibnamefont {Vuletić}},\ and\ \bibinfo {author} {\bibfnamefont {M.~D.}\ \bibnamefont {Lukin}},\ }\bibfield  {title} {\bibinfo {title} {Continuous operation of a coherent 3,000-qubit system},\ }\href@noop {} {\bibfield  {journal} {\bibinfo  {journal} {Nature}\ }\textbf {\bibinfo {volume} {646}},\ \bibinfo {pages} {1075} (\bibinfo {year} {2025})}\BibitemShut {NoStop}%
\bibitem [{\citenamefont {Lin}\ \emph {et~al.}(2025)\citenamefont {Lin}, \citenamefont {Zhong}, \citenamefont {Li}, \citenamefont {Zhao}, \citenamefont {Zheng}, \citenamefont {Hu}, \citenamefont {Wu}, \citenamefont {Wu}, \citenamefont {Ma}, \citenamefont {Gao}, \citenamefont {Zhu}, \citenamefont {Su}, \citenamefont {Ouyang}, \citenamefont {Zhang}, \citenamefont {Rui}, \citenamefont {Chen}, \citenamefont {Lu},\ and\ \citenamefont {Pan}}]{lin2025ai}%
  \BibitemOpen
  \bibfield  {author} {\bibinfo {author} {\bibfnamefont {R.}~\bibnamefont {Lin}}, \bibinfo {author} {\bibfnamefont {H.-S.}\ \bibnamefont {Zhong}}, \bibinfo {author} {\bibfnamefont {Y.}~\bibnamefont {Li}}, \bibinfo {author} {\bibfnamefont {Z.-R.}\ \bibnamefont {Zhao}}, \bibinfo {author} {\bibfnamefont {L.-T.}\ \bibnamefont {Zheng}}, \bibinfo {author} {\bibfnamefont {T.-R.}\ \bibnamefont {Hu}}, \bibinfo {author} {\bibfnamefont {H.-M.}\ \bibnamefont {Wu}}, \bibinfo {author} {\bibfnamefont {Z.}~\bibnamefont {Wu}}, \bibinfo {author} {\bibfnamefont {W.-J.}\ \bibnamefont {Ma}}, \bibinfo {author} {\bibfnamefont {Y.}~\bibnamefont {Gao}}, \bibinfo {author} {\bibfnamefont {Y.-K.}\ \bibnamefont {Zhu}}, \bibinfo {author} {\bibfnamefont {Z.-F.}\ \bibnamefont {Su}}, \bibinfo {author} {\bibfnamefont {W.-L.}\ \bibnamefont {Ouyang}}, \bibinfo {author} {\bibfnamefont {Y.-C.}\ \bibnamefont {Zhang}}, \bibinfo {author} {\bibfnamefont {J.}~\bibnamefont {Rui}}, \bibinfo {author} {\bibfnamefont {M.-C.}\ \bibnamefont {Chen}}, \bibinfo {author} {\bibfnamefont {C.-Y.}\ \bibnamefont {Lu}},\ and\ \bibinfo {author} {\bibfnamefont {J.-W.}\ \bibnamefont {Pan}},\ }\bibfield  {title} {\bibinfo {title} {{AI}-enabled parallel assembly of thousands of defect-free neutral atom arrays},\ }\href@noop {} {\bibfield  {journal} {\bibinfo  {journal} {Physical Review Letters}\ }\textbf {\bibinfo {volume} {135}},\ \bibinfo {pages} {060602} (\bibinfo {year} {2025})}\BibitemShut {NoStop}%
\bibitem [{\citenamefont {Saffman}(2019)}]{saffman2019quantum}%
  \BibitemOpen
  \bibfield  {author} {\bibinfo {author} {\bibfnamefont {M.}~\bibnamefont {Saffman}},\ }\bibfield  {title} {\bibinfo {title} {Quantum computing with neutral atoms},\ }\href@noop {} {\bibfield  {journal} {\bibinfo  {journal} {National Science Review}\ }\textbf {\bibinfo {volume} {6}},\ \bibinfo {pages} {24} (\bibinfo {year} {2019})}\BibitemShut {NoStop}%
\bibitem [{\citenamefont {Beverland}\ \emph {et~al.}(2022)\citenamefont {Beverland}, \citenamefont {Murali}, \citenamefont {Troyer}, \citenamefont {Svore}, \citenamefont {Hoefler}, \citenamefont {Kliuchnikov}, \citenamefont {Low}, \citenamefont {Soeken}, \citenamefont {Sundaram},\ and\ \citenamefont {Vaschillo}}]{beverland2022assessing}%
  \BibitemOpen
  \bibfield  {author} {\bibinfo {author} {\bibfnamefont {M.~E.}\ \bibnamefont {Beverland}}, \bibinfo {author} {\bibfnamefont {P.}~\bibnamefont {Murali}}, \bibinfo {author} {\bibfnamefont {M.}~\bibnamefont {Troyer}}, \bibinfo {author} {\bibfnamefont {K.~M.}\ \bibnamefont {Svore}}, \bibinfo {author} {\bibfnamefont {T.}~\bibnamefont {Hoefler}}, \bibinfo {author} {\bibfnamefont {V.}~\bibnamefont {Kliuchnikov}}, \bibinfo {author} {\bibfnamefont {G.~H.}\ \bibnamefont {Low}}, \bibinfo {author} {\bibfnamefont {M.}~\bibnamefont {Soeken}}, \bibinfo {author} {\bibfnamefont {A.}~\bibnamefont {Sundaram}},\ and\ \bibinfo {author} {\bibfnamefont {A.}~\bibnamefont {Vaschillo}},\ }\bibfield  {title} {\bibinfo {title} {Assessing requirements to scale to practical quantum advantage},\ }\href@noop {} {\bibfield  {journal} {\bibinfo  {journal} {arXiv preprint arXiv:2211.07629}\ } (\bibinfo {year} {2022})}\BibitemShut {NoStop}%
\bibitem [{\citenamefont {Cain}\ \emph {et~al.}(2026)\citenamefont {Cain}, \citenamefont {Xu}, \citenamefont {King}, \citenamefont {Picard}, \citenamefont {Levine}, \citenamefont {Endres}, \citenamefont {Preskill}, \citenamefont {Huang},\ and\ \citenamefont {Bluvstein}}]{cain2026shor}%
  \BibitemOpen
  \bibfield  {author} {\bibinfo {author} {\bibfnamefont {M.}~\bibnamefont {Cain}}, \bibinfo {author} {\bibfnamefont {Q.}~\bibnamefont {Xu}}, \bibinfo {author} {\bibfnamefont {R.}~\bibnamefont {King}}, \bibinfo {author} {\bibfnamefont {L.~R.}\ \bibnamefont {Picard}}, \bibinfo {author} {\bibfnamefont {H.}~\bibnamefont {Levine}}, \bibinfo {author} {\bibfnamefont {M.}~\bibnamefont {Endres}}, \bibinfo {author} {\bibfnamefont {J.}~\bibnamefont {Preskill}}, \bibinfo {author} {\bibfnamefont {H.-Y.}\ \bibnamefont {Huang}},\ and\ \bibinfo {author} {\bibfnamefont {D.}~\bibnamefont {Bluvstein}},\ }\bibfield  {title} {\bibinfo {title} {Shor's algorithm is possible with as few as 10,000 reconfigurable atomic qubits},\ }\href@noop {} {\bibfield  {journal} {\bibinfo  {journal} {arXiv preprint arXiv:2603.28627}\ } (\bibinfo {year} {2026})}\BibitemShut {NoStop}%
\bibitem [{\citenamefont {Barredo}\ \emph {et~al.}(2016)\citenamefont {Barredo}, \citenamefont {De~L{\'e}s{\'e}leuc}, \citenamefont {Lienhard}, \citenamefont {Lahaye},\ and\ \citenamefont {Browaeys}}]{barredo2016atom}%
  \BibitemOpen
  \bibfield  {author} {\bibinfo {author} {\bibfnamefont {D.}~\bibnamefont {Barredo}}, \bibinfo {author} {\bibfnamefont {S.}~\bibnamefont {De~L{\'e}s{\'e}leuc}}, \bibinfo {author} {\bibfnamefont {V.}~\bibnamefont {Lienhard}}, \bibinfo {author} {\bibfnamefont {T.}~\bibnamefont {Lahaye}},\ and\ \bibinfo {author} {\bibfnamefont {A.}~\bibnamefont {Browaeys}},\ }\bibfield  {title} {\bibinfo {title} {An atom-by-atom assembler of defect-free arbitrary two-dimensional atomic arrays},\ }\href@noop {} {\bibfield  {journal} {\bibinfo  {journal} {Science}\ }\textbf {\bibinfo {volume} {354}},\ \bibinfo {pages} {1021} (\bibinfo {year} {2016})}\BibitemShut {NoStop}%
\bibitem [{\citenamefont {Endres}\ \emph {et~al.}(2016)\citenamefont {Endres}, \citenamefont {Bernien}, \citenamefont {Keesling}, \citenamefont {Levine}, \citenamefont {Anschuetz}, \citenamefont {Krajenbrink}, \citenamefont {Senko}, \citenamefont {Vuletic}, \citenamefont {Greiner},\ and\ \citenamefont {Lukin}}]{endres2016atom}%
  \BibitemOpen
  \bibfield  {author} {\bibinfo {author} {\bibfnamefont {M.}~\bibnamefont {Endres}}, \bibinfo {author} {\bibfnamefont {H.}~\bibnamefont {Bernien}}, \bibinfo {author} {\bibfnamefont {A.}~\bibnamefont {Keesling}}, \bibinfo {author} {\bibfnamefont {H.}~\bibnamefont {Levine}}, \bibinfo {author} {\bibfnamefont {E.~R.}\ \bibnamefont {Anschuetz}}, \bibinfo {author} {\bibfnamefont {A.}~\bibnamefont {Krajenbrink}}, \bibinfo {author} {\bibfnamefont {C.}~\bibnamefont {Senko}}, \bibinfo {author} {\bibfnamefont {V.}~\bibnamefont {Vuletic}}, \bibinfo {author} {\bibfnamefont {M.}~\bibnamefont {Greiner}},\ and\ \bibinfo {author} {\bibfnamefont {M.~D.}\ \bibnamefont {Lukin}},\ }\bibfield  {title} {\bibinfo {title} {Atom-by-atom assembly of defect-free one-dimensional cold atom arrays},\ }\href@noop {} {\bibfield  {journal} {\bibinfo  {journal} {Science}\ }\textbf {\bibinfo {volume} {354}},\ \bibinfo {pages} {1024} (\bibinfo {year} {2016})}\BibitemShut {NoStop}%
\bibitem [{\citenamefont {Schymik}\ \emph {et~al.}(2020)\citenamefont {Schymik}, \citenamefont {Lienhard}, \citenamefont {Barredo}, \citenamefont {Scholl}, \citenamefont {Williams}, \citenamefont {Browaeys},\ and\ \citenamefont {Lahaye}}]{schymik2020enhanced}%
  \BibitemOpen
  \bibfield  {author} {\bibinfo {author} {\bibfnamefont {K.-N.}\ \bibnamefont {Schymik}}, \bibinfo {author} {\bibfnamefont {V.}~\bibnamefont {Lienhard}}, \bibinfo {author} {\bibfnamefont {D.}~\bibnamefont {Barredo}}, \bibinfo {author} {\bibfnamefont {P.}~\bibnamefont {Scholl}}, \bibinfo {author} {\bibfnamefont {H.}~\bibnamefont {Williams}}, \bibinfo {author} {\bibfnamefont {A.}~\bibnamefont {Browaeys}},\ and\ \bibinfo {author} {\bibfnamefont {T.}~\bibnamefont {Lahaye}},\ }\bibfield  {title} {\bibinfo {title} {Enhanced atom-by-atom assembly of arbitrary tweezer arrays},\ }\href@noop {} {\bibfield  {journal} {\bibinfo  {journal} {Physical Review A}\ }\textbf {\bibinfo {volume} {102}},\ \bibinfo {pages} {063107} (\bibinfo {year} {2020})}\BibitemShut {NoStop}%
\bibitem [{\citenamefont {Tian}\ \emph {et~al.}(2023)\citenamefont {Tian}, \citenamefont {Wee}, \citenamefont {Qu}, \citenamefont {Lim}, \citenamefont {Datla}, \citenamefont {Koh},\ and\ \citenamefont {Loh}}]{tian2023parallel}%
  \BibitemOpen
  \bibfield  {author} {\bibinfo {author} {\bibfnamefont {W.}~\bibnamefont {Tian}}, \bibinfo {author} {\bibfnamefont {W.~J.}\ \bibnamefont {Wee}}, \bibinfo {author} {\bibfnamefont {A.}~\bibnamefont {Qu}}, \bibinfo {author} {\bibfnamefont {B.~J.~M.}\ \bibnamefont {Lim}}, \bibinfo {author} {\bibfnamefont {P.~R.}\ \bibnamefont {Datla}}, \bibinfo {author} {\bibfnamefont {V.~P.~W.}\ \bibnamefont {Koh}},\ and\ \bibinfo {author} {\bibfnamefont {H.}~\bibnamefont {Loh}},\ }\bibfield  {title} {\bibinfo {title} {Parallel assembly of arbitrary defect-free atom arrays with a multitweezer algorithm},\ }\href@noop {} {\bibfield  {journal} {\bibinfo  {journal} {Physical Review Applied}\ }\textbf {\bibinfo {volume} {19}},\ \bibinfo {pages} {034048} (\bibinfo {year} {2023})}\BibitemShut {NoStop}%
\bibitem [{\citenamefont {Wang}\ \emph {et~al.}(2023)\citenamefont {Wang}, \citenamefont {Zhang}, \citenamefont {Zhang}, \citenamefont {Mei}, \citenamefont {Wang}, \citenamefont {Hu},\ and\ \citenamefont {Chen}}]{wang2023accelerating}%
  \BibitemOpen
  \bibfield  {author} {\bibinfo {author} {\bibfnamefont {S.}~\bibnamefont {Wang}}, \bibinfo {author} {\bibfnamefont {W.}~\bibnamefont {Zhang}}, \bibinfo {author} {\bibfnamefont {T.}~\bibnamefont {Zhang}}, \bibinfo {author} {\bibfnamefont {S.}~\bibnamefont {Mei}}, \bibinfo {author} {\bibfnamefont {Y.}~\bibnamefont {Wang}}, \bibinfo {author} {\bibfnamefont {J.}~\bibnamefont {Hu}},\ and\ \bibinfo {author} {\bibfnamefont {W.}~\bibnamefont {Chen}},\ }\bibfield  {title} {\bibinfo {title} {Accelerating the assembly of defect-free atomic arrays with maximum parallelisms},\ }\href@noop {} {\bibfield  {journal} {\bibinfo  {journal} {Physical Review Applied}\ }\textbf {\bibinfo {volume} {19}},\ \bibinfo {pages} {054032} (\bibinfo {year} {2023})}\BibitemShut {NoStop}%
\bibitem [{\citenamefont {Kim}\ \emph {et~al.}(2016)\citenamefont {Kim}, \citenamefont {Lee}, \citenamefont {Lee}, \citenamefont {Jo}, \citenamefont {Song},\ and\ \citenamefont {Ahn}}]{kim2016situ}%
  \BibitemOpen
  \bibfield  {author} {\bibinfo {author} {\bibfnamefont {H.}~\bibnamefont {Kim}}, \bibinfo {author} {\bibfnamefont {W.}~\bibnamefont {Lee}}, \bibinfo {author} {\bibfnamefont {H.-g.}\ \bibnamefont {Lee}}, \bibinfo {author} {\bibfnamefont {H.}~\bibnamefont {Jo}}, \bibinfo {author} {\bibfnamefont {Y.}~\bibnamefont {Song}},\ and\ \bibinfo {author} {\bibfnamefont {J.}~\bibnamefont {Ahn}},\ }\bibfield  {title} {\bibinfo {title} {In situ single-atom array synthesis using dynamic holographic optical tweezers},\ }\href@noop {} {\bibfield  {journal} {\bibinfo  {journal} {{Nature Communications}}\ }\textbf {\bibinfo {volume} {7}},\ \bibinfo {pages} {13317} (\bibinfo {year} {2016})}\BibitemShut {NoStop}%
\bibitem [{\citenamefont {Kim}\ \emph {et~al.}(2019)\citenamefont {Kim}, \citenamefont {Kim}, \citenamefont {Lee},\ and\ \citenamefont {Ahn}}]{kim2019gerchberg}%
  \BibitemOpen
  \bibfield  {author} {\bibinfo {author} {\bibfnamefont {H.}~\bibnamefont {Kim}}, \bibinfo {author} {\bibfnamefont {M.}~\bibnamefont {Kim}}, \bibinfo {author} {\bibfnamefont {W.}~\bibnamefont {Lee}},\ and\ \bibinfo {author} {\bibfnamefont {J.}~\bibnamefont {Ahn}},\ }\bibfield  {title} {\bibinfo {title} {Gerchberg-saxton algorithm for fast and efficient atom rearrangement in optical tweezer traps},\ }\href@noop {} {\bibfield  {journal} {\bibinfo  {journal} {{Optics Express}}\ }\textbf {\bibinfo {volume} {27}},\ \bibinfo {pages} {2184} (\bibinfo {year} {2019})}\BibitemShut {NoStop}%
\bibitem [{\citenamefont {Lee}\ \emph {et~al.}(2017)\citenamefont {Lee}, \citenamefont {Kim},\ and\ \citenamefont {Ahn}}]{lee2017defect}%
  \BibitemOpen
  \bibfield  {author} {\bibinfo {author} {\bibfnamefont {W.}~\bibnamefont {Lee}}, \bibinfo {author} {\bibfnamefont {H.}~\bibnamefont {Kim}},\ and\ \bibinfo {author} {\bibfnamefont {J.}~\bibnamefont {Ahn}},\ }\bibfield  {title} {\bibinfo {title} {Defect-free atomic array formation using the hungarian matching algorithm},\ }\href@noop {} {\bibfield  {journal} {\bibinfo  {journal} {Physical Review A}\ }\textbf {\bibinfo {volume} {95}},\ \bibinfo {pages} {053424} (\bibinfo {year} {2017})}\BibitemShut {NoStop}%
\bibitem [{\citenamefont {Manetsch}\ \emph {et~al.}(2025)\citenamefont {Manetsch}, \citenamefont {Nomura}, \citenamefont {Bataille}, \citenamefont {Lv}, \citenamefont {Leung},\ and\ \citenamefont {Endres}}]{manetsch2025tweezer}%
  \BibitemOpen
  \bibfield  {author} {\bibinfo {author} {\bibfnamefont {H.~J.}\ \bibnamefont {Manetsch}}, \bibinfo {author} {\bibfnamefont {G.}~\bibnamefont {Nomura}}, \bibinfo {author} {\bibfnamefont {E.}~\bibnamefont {Bataille}}, \bibinfo {author} {\bibfnamefont {X.}~\bibnamefont {Lv}}, \bibinfo {author} {\bibfnamefont {K.~H.}\ \bibnamefont {Leung}},\ and\ \bibinfo {author} {\bibfnamefont {M.}~\bibnamefont {Endres}},\ }\bibfield  {title} {\bibinfo {title} {A tweezer array with 6,100 highly coherent atomic qubits},\ }\href@noop {} {\bibfield  {journal} {\bibinfo  {journal} {Nature}\ }\textbf {\bibinfo {volume} {647}},\ \bibinfo {pages} {60} (\bibinfo {year} {2025})}\BibitemShut {NoStop}%
\bibitem [{\citenamefont {Khalil}\ \emph {et~al.}(2017)\citenamefont {Khalil}, \citenamefont {Dai}, \citenamefont {Zhang}, \citenamefont {Dilkina},\ and\ \citenamefont {Song}}]{khalil2017learning}%
  \BibitemOpen
  \bibfield  {author} {\bibinfo {author} {\bibfnamefont {E.}~\bibnamefont {Khalil}}, \bibinfo {author} {\bibfnamefont {H.}~\bibnamefont {Dai}}, \bibinfo {author} {\bibfnamefont {Y.}~\bibnamefont {Zhang}}, \bibinfo {author} {\bibfnamefont {B.}~\bibnamefont {Dilkina}},\ and\ \bibinfo {author} {\bibfnamefont {L.}~\bibnamefont {Song}},\ }\bibfield  {title} {\bibinfo {title} {Learning combinatorial optimization algorithms over graphs},\ }\href@noop {} {\bibfield  {journal} {\bibinfo  {journal} {{Advances in Neural Information Processing Systems}}\ }\textbf {\bibinfo {volume} {30}} (\bibinfo {year} {2017})}\BibitemShut {NoStop}%
\bibitem [{\citenamefont {Gerchberg}(1972)}]{gerchberg1972practical}%
  \BibitemOpen
  \bibfield  {author} {\bibinfo {author} {\bibfnamefont {R.~W.}\ \bibnamefont {Gerchberg}},\ }\bibfield  {title} {\bibinfo {title} {A practical algorithm for the determination of the phase from image and diffraction plane pictures},\ }\href@noop {} {\bibfield  {journal} {\bibinfo  {journal} {Optik}\ }\textbf {\bibinfo {volume} {35}},\ \bibinfo {pages} {237} (\bibinfo {year} {1972})}\BibitemShut {NoStop}%
\bibitem [{\citenamefont {Di~Leonardo}\ \emph {et~al.}(2007)\citenamefont {Di~Leonardo}, \citenamefont {Ianni},\ and\ \citenamefont {Ruocco}}]{di2007computer}%
  \BibitemOpen
  \bibfield  {author} {\bibinfo {author} {\bibfnamefont {R.}~\bibnamefont {Di~Leonardo}}, \bibinfo {author} {\bibfnamefont {F.}~\bibnamefont {Ianni}},\ and\ \bibinfo {author} {\bibfnamefont {G.}~\bibnamefont {Ruocco}},\ }\bibfield  {title} {\bibinfo {title} {Computer generation of optimal holograms for optical trap arrays},\ }\href@noop {} {\bibfield  {journal} {\bibinfo  {journal} {{Optics Express}}\ }\textbf {\bibinfo {volume} {15}},\ \bibinfo {pages} {1913} (\bibinfo {year} {2007})}\BibitemShut {NoStop}%
\bibitem [{\citenamefont {Kuhn}(1955)}]{kuhn1955hungarian}%
  \BibitemOpen
  \bibfield  {author} {\bibinfo {author} {\bibfnamefont {H.~W.}\ \bibnamefont {Kuhn}},\ }\bibfield  {title} {\bibinfo {title} {The hungarian method for the assignment problem},\ }\href@noop {} {\bibfield  {journal} {\bibinfo  {journal} {{Naval Research Logistics Quarterly}}\ }\textbf {\bibinfo {volume} {2}},\ \bibinfo {pages} {83} (\bibinfo {year} {1955})}\BibitemShut {NoStop}%
\bibitem [{\citenamefont {Jonker}\ and\ \citenamefont {Volgenant}(1987)}]{jonker1987shortest}%
  \BibitemOpen
  \bibfield  {author} {\bibinfo {author} {\bibfnamefont {R.}~\bibnamefont {Jonker}}\ and\ \bibinfo {author} {\bibfnamefont {A.}~\bibnamefont {Volgenant}},\ }\bibfield  {title} {\bibinfo {title} {A shortest augmenting path algorithm for dense and sparse linear assignment problems},\ }\href@noop {} {\bibfield  {journal} {\bibinfo  {journal} {Computing}\ }\textbf {\bibinfo {volume} {38}},\ \bibinfo {pages} {325} (\bibinfo {year} {1987})}\BibitemShut {NoStop}%
\bibitem [{\citenamefont {Bertsekas}(1988)}]{bertsekas1988auction}%
  \BibitemOpen
  \bibfield  {author} {\bibinfo {author} {\bibfnamefont {D.~P.}\ \bibnamefont {Bertsekas}},\ }\bibfield  {title} {\bibinfo {title} {The auction algorithm: A distributed relaxation method for the assignment problem},\ }\href@noop {} {\bibfield  {journal} {\bibinfo  {journal} {{Annals of Operations Research}}\ }\textbf {\bibinfo {volume} {14}},\ \bibinfo {pages} {105} (\bibinfo {year} {1988})}\BibitemShut {NoStop}%
\bibitem [{\citenamefont {Pasienski}\ and\ \citenamefont {DeMarco}(2008)}]{pasienski2008high}%
  \BibitemOpen
  \bibfield  {author} {\bibinfo {author} {\bibfnamefont {M.}~\bibnamefont {Pasienski}}\ and\ \bibinfo {author} {\bibfnamefont {B.}~\bibnamefont {DeMarco}},\ }\bibfield  {title} {\bibinfo {title} {A high-accuracy algorithm for designing arbitrary holographic atom traps},\ }\href@noop {} {\bibfield  {journal} {\bibinfo  {journal} {{Optics Express}}\ }\textbf {\bibinfo {volume} {16}},\ \bibinfo {pages} {2176} (\bibinfo {year} {2008})}\BibitemShut {NoStop}%
\bibitem [{Note1()}]{Note1}%
  \BibitemOpen
  \bibinfo {note} {In practice, the initial stochastic loading typically captures more atoms than the required target array size. To assemble a defect-free array, the redundant atoms must be released. In our algorithm, before switching off the redundant optical tweezers at the fourth frame, we apply randomized spatial perturbations to their positions during the first three frames to avoid optical interference. Because the SLM hologram generates all tweezers simultaneously via global Fourier interference, the random displacement of these redundant tweezers introduces global potential perturbations, leading to the slightly larger phase and intensity fluctuations observed in these initial frames. Once the redundant tweezers are completely removed at the fourth frame, the remaining tweezers move strictly along smooth, deterministic linear trajectories, allowing the optical potential to stabilize rapidly.}\BibitemShut {Stop}%
\bibitem [{\citenamefont {Wei}\ \emph {et~al.}(2026)\citenamefont {Wei}, \citenamefont {Li}, \citenamefont {Karve}, \citenamefont {Shaw}, \citenamefont {Schuster},\ and\ \citenamefont {Simon}}]{wei202610}%
  \BibitemOpen
  \bibfield  {author} {\bibinfo {author} {\bibfnamefont {X.}~\bibnamefont {Wei}}, \bibinfo {author} {\bibfnamefont {Z.}~\bibnamefont {Li}}, \bibinfo {author} {\bibfnamefont {A.~V.}\ \bibnamefont {Karve}}, \bibinfo {author} {\bibfnamefont {A.~L.}\ \bibnamefont {Shaw}}, \bibinfo {author} {\bibfnamefont {D.~I.}\ \bibnamefont {Schuster}},\ and\ \bibinfo {author} {\bibfnamefont {J.}~\bibnamefont {Simon}},\ }\bibfield  {title} {\bibinfo {title} {A 10 megahertz spatial light modulator},\ }\href@noop {} {\bibfield  {journal} {\bibinfo  {journal} {arXiv preprint arXiv:2601.08906}\ } (\bibinfo {year} {2026})}\BibitemShut {NoStop}%
\bibitem [{\citenamefont {Bytyqi}\ \emph {et~al.}(2026)\citenamefont {Bytyqi}, \citenamefont {Sinclair}, \citenamefont {Ramette},\ and\ \citenamefont {Vuleti{\'c}}}]{bytyqi2026device}%
  \BibitemOpen
  \bibfield  {author} {\bibinfo {author} {\bibfnamefont {E.}~\bibnamefont {Bytyqi}}, \bibinfo {author} {\bibfnamefont {J.}~\bibnamefont {Sinclair}}, \bibinfo {author} {\bibfnamefont {J.}~\bibnamefont {Ramette}},\ and\ \bibinfo {author} {\bibfnamefont {V.}~\bibnamefont {Vuleti{\'c}}},\ }\bibfield  {title} {\bibinfo {title} {Device for mhz-rate rastering of arbitrary 2d optical potentials},\ }\href@noop {} {\bibfield  {journal} {\bibinfo  {journal} {arXiv preprint arXiv:2602.16025}\ } (\bibinfo {year} {2026})}\BibitemShut {NoStop}%
\bibitem [{\citenamefont {Breuckmann}\ and\ \citenamefont {Eberhardt}(2021)}]{breuckmann2021quantum}%
  \BibitemOpen
  \bibfield  {author} {\bibinfo {author} {\bibfnamefont {N.~P.}\ \bibnamefont {Breuckmann}}\ and\ \bibinfo {author} {\bibfnamefont {J.~N.}\ \bibnamefont {Eberhardt}},\ }\bibfield  {title} {\bibinfo {title} {Quantum low-density parity-check codes},\ }\href@noop {} {\bibfield  {journal} {\bibinfo  {journal} {PRX quantum}\ }\textbf {\bibinfo {volume} {2}},\ \bibinfo {pages} {040101} (\bibinfo {year} {2021})}\BibitemShut {NoStop}%
\bibitem [{\citenamefont {Xu}\ \emph {et~al.}(2024)\citenamefont {Xu}, \citenamefont {Bonilla~Ataides}, \citenamefont {Pattison}, \citenamefont {Raveendran}, \citenamefont {Bluvstein}, \citenamefont {Wurtz}, \citenamefont {Vasi{\'c}}, \citenamefont {Lukin}, \citenamefont {Jiang},\ and\ \citenamefont {Zhou}}]{xu2024constant}%
  \BibitemOpen
  \bibfield  {author} {\bibinfo {author} {\bibfnamefont {Q.}~\bibnamefont {Xu}}, \bibinfo {author} {\bibfnamefont {J.~P.}\ \bibnamefont {Bonilla~Ataides}}, \bibinfo {author} {\bibfnamefont {C.~A.}\ \bibnamefont {Pattison}}, \bibinfo {author} {\bibfnamefont {N.}~\bibnamefont {Raveendran}}, \bibinfo {author} {\bibfnamefont {D.}~\bibnamefont {Bluvstein}}, \bibinfo {author} {\bibfnamefont {J.}~\bibnamefont {Wurtz}}, \bibinfo {author} {\bibfnamefont {B.}~\bibnamefont {Vasi{\'c}}}, \bibinfo {author} {\bibfnamefont {M.~D.}\ \bibnamefont {Lukin}}, \bibinfo {author} {\bibfnamefont {L.}~\bibnamefont {Jiang}},\ and\ \bibinfo {author} {\bibfnamefont {H.}~\bibnamefont {Zhou}},\ }\bibfield  {title} {\bibinfo {title} {Constant-overhead fault-tolerant quantum computation with reconfigurable atom arrays},\ }\href@noop {} {\bibfield  {journal} {\bibinfo  {journal} {Nature Physics}\ }\textbf {\bibinfo {volume} {20}},\ \bibinfo {pages} {1084} (\bibinfo {year} {2024})}\BibitemShut {NoStop}%
\bibitem [{Note2()}]{Note2}%
  \BibitemOpen
  \bibinfo {note} {``Zhuifeng,'' which literally translates to ``Chasing the Wind,'' was the fastest and most beloved horse of Emperor Qin Shi Huang, the first emperor of the Qin Dynasty.}\BibitemShut {Stop}%
\end{thebibliography}%

\end{document}